\documentclass[%
 reprint,
superscriptaddress,
 amsmath,amssymb,
 aps,
]{revtex4-2}

\usepackage{braket}
\usepackage{booktabs}
\usepackage{placeins}
\usepackage{graphicx}
\usepackage{dcolumn}
\usepackage{bm}
\usepackage[dvipsnames]{xcolor}
\usepackage{siunitx}
\DeclareSIUnit\gauss{G}
\usepackage[export]{adjustbox}
\usepackage{comment}
\usepackage{wasysym}

\begin{document}

\title{Trapped Ion Quantum Networking and Telecommunications Coexisting on One Fiber}

\author{Denton Wu}
\email{denton.wu@duke.edu}
    \affiliation{%
 Duke Quantum Center, Duke University, Durham, NC 27708
}%
\author{Mingzhe Han}
\affiliation{%
 Department of Electrical and Computer Engineering, Duke University, Durham, NC 27708
}%

\author{Zehao Wang}
\affiliation{%
 Department of Electrical and Computer Engineering, Duke University, Durham, NC 27708
}%
\author{Ana Luiza Ferrari}
\affiliation{%
 Duke Quantum Center, Duke University, Durham, NC 27708
}%

\author{Mika A. Zalewski}
\affiliation{%
 Duke Quantum Center, Duke University, Durham, NC 27708
}%

\author{Yuanheng Xie}
\affiliation{%
 Duke Quantum Center, Duke University, Durham, NC 27708
}%
\affiliation{Joint Quantum Institute and Department of Physics, University of Maryland, College Park, MD 20742, USA}
\affiliation{National Quantum Laboratory (QLab), University of Maryland, College Park, MD 20742, USA}

\author{Tingjun Chen}
\affiliation{%
 Department of Electrical and Computer Engineering, Duke University, Durham, NC 27708
}%

\author{Norbert M. Linke}
\email{linke@umd.edu}
\affiliation{%
 Duke Quantum Center, Duke University, Durham, NC 27708
}%
\affiliation{Joint Quantum Institute and Department of Physics, University of Maryland, College Park, MD 20742, USA}
\affiliation{National Quantum Laboratory (QLab), University of Maryland, College Park, MD 20742, USA}

\date{\today}

\begin{abstract}

Research into long-distance quantum memory-based networking to date has exclusively used dark fibers. This avoids the detector background from telecommunications (telecom) traffic, but as a result excludes many fibers deployed in the field. If memory-photon entanglement and telecom signals coexist on one fiber, the entire classical fiber infrastructure becomes available for quantum links. We present the first experimental demonstration of such coexistence. Ion-photon entanglement using \SI{1092}{\nano\meter} photons emitted by a \textsuperscript{88}Sr\textsuperscript{+} ion is distributed over a deployed \SI{2.8}{\kilo\meter} fiber loop which carries Ethernet and 5G traffic. All classical control signals required to coordinate the quantum transmitter and receiver systems co-propagate on the same fiber. These include fiber sensing for polarization stabilization. Our results demonstrate that memory-based quantum networks can be realized on active classical network infrastructure.

\end{abstract}

\maketitle

\section{\label{sec:intro}Introduction}

\begin{figure}[b]
    \centering
    \includegraphics[width=0.75\linewidth]{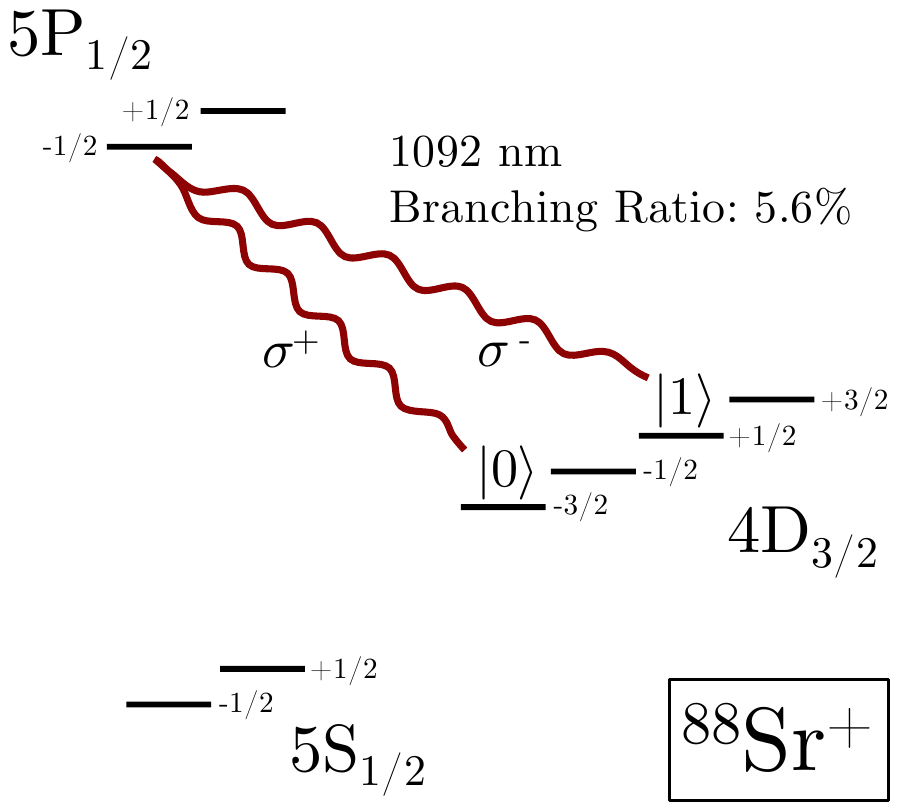}
    \caption{Entanglement generation scheme. A \textsuperscript{88}Sr\textsuperscript{+} ion is excited to the $\ket{5P_{1/2}, -1/2}$ state. Decay along the \SI{1092}{\nano\meter} transition to the $4D_{3/2}$ level occurs with 5.6\% branching ratio. Light collection along the magnetic field yields a photon in a $\sigma^{\pm}$ polarization superposition, entangled with the ion qubit states $\ket{4D_{3/2}, -3/2}$ and $\ket{4D_{3/2}, +1/2}$.}
    \label{fig:levelDiagram}
\end{figure}

Practical quantum computing applications such as Shor's algorithm, Grover's search algorithm, and quantum simulation necessitate systems with orders of magnitude more qubits than currently available~\cite{gidney2021factor,grassl2016applying,lee2021even}. Quantum memory-based platforms such as trapped ions, neutral atoms, superconducting circuits, and solid-state qubits face limitations on the size of a single processing module~\cite{pino2021demonstration,HowToWire,menssen2026strategicplanneutralatom,wang2026neutralatomquantumcomputing,croot2025enabling,de2021materials}. Photonically networking modules together offers a path for scaling beyond these limits and enables additional applications such as sensing and blind computing~\cite{ChrisPRA,OxfordDistributed,OxfordBlind,FitzsimmonsBlind,komar2014quantum,OxfordClocks}. This requires the ability to distribute high-fidelity memory-photon entanglement between modules. Trapped ions, neutral atoms, and solid-state systems have demonstrated networking results toward this end, with ions achieving the highest rate and fidelity~\cite{Jameson,van2022entangling,liu2024creation,ritter2012elementary,knaut2024entanglement,Hansen2024, bernien2013heralded}. For meter-scale networking with ions, visible photons can be used~\cite{SagnikTimeBin,StephensonHighRate}. Converting photons to telecommunication (telecom) wavelengths can extend networking ranges to the kilometer scale for long distance functionality connecting buildings within a city~\cite{LanyonMultimode,Duan12km,Kucera14,Qudsia}. 

Networking at these distances must be implemented using municipal fiber resources. These fibers are frequently occupied by wavelength-division multiplexed telecom traffic, such as Ethernet, 5G, and fiber sensing services, leaving limited dark fibers for quantum links~\cite{wang2023field,wang2025toward}. Multiplexing single photons onto fibers carrying telecom traffic offers a means to utilize more than just the dark fibers, boosting the number of quantum links attainable on the network. This has been explored with purely photonic entanglement distribution, but never before with memory-photon entanglement distribution~\cite{OpticaCoexistence,yuan2019quantum,rahmouni2024100,sena2025high,thomas2023designing}. Crucial to this approach is the suppression of detector background to levels below photon counts. Employing frequency conversion to obtain photon wavelengths within telecom bands often precludes coexistence with telecom traffic as filtering traffic from the single photons becomes challenging. Telecom signals undergo spontaneous Raman scattering with phonons in fiber, which induces photon noise around signal wavelengths that can cover tens to hundreds of nanometers of bandwidth~\cite{OpticaCoexistence,Agrawal}. Even with narrow filters, this noise can leak into detectors.

There exist transition wavelengths that lie outside the noise band which are capable of traversing city-scale distances without frequency conversion~\cite{zalewski2026kilometer,Lanyon230,Covey}. These alleviate filter leakage and enable multiplexing photons with telecom traffic. Moreover, using such transitions circumvents the fidelity and experimental overhead costs of frequency conversion. The tradeoff is higher fiber attenuation. In this work, we use the \SI{1092}{\nano\meter} transition in \textsuperscript{88}Sr\textsuperscript{+} to distribute entanglement. \SI{1092}{\nano\meter} has a loss rate of 0.7 dB/km in fiber, higher than the optimal telecom wavelength loss of 0.18 dB/km. This must be weighed against the gains in quantum link number enabled by multiplexing. We show the usability range of our scheme to be several kilometers, offering a tool for city-scale networking applications that require high quantum link capacity through multiplexing.

Building on previous work~\cite{zalewski2026kilometer,wang2026coexistence}, we demonstrate quantum-classical coexistence by distributing entanglement between a \textsuperscript{88}Sr\textsuperscript{+} ion and the polarization of \SI{1092}{\nano\meter} photons (Figure~\ref{fig:levelDiagram}) over a deployed fiber carrying telecom signals. High fidelity is achieved for over 12 hours, all while high-speed Ethernet and 5G radio-over-fiber traffic traverse the fiber at \SI{1549.32}{\nano\meter} and \SI{1542.93}{\nano\meter}, respectively. Deployed fibers can experience thermal and mechanical instabilities that cause polarization drift, reducing state fidelity if left unchecked~\cite{BARQNET,Kucera14}. We implement polarization drift sensing over the fiber at \SI{1530}{\nano\meter} during entanglement distribution. This triggers polarization drift correction if instability is detected. All distributed control signaling required to generate the entanglement is implemented on the fiber at telecom wavelengths between \SI{1470}{\nano\meter} and \SI{1610}{\nano\meter}. Wavelength-division multiplexers (WDMs), interference filters, and the chromatic dispersion of free-space optics before the photon detector are sufficient to reduce telecom-induced detector background counts to negligible levels. Our results show that selecting a photon wavelength outside telecom bands enables coexistence with telecom signals, facilitating quantum link deployment over fibers already in use. This paves a path for realizing memory-based quantum networks on the city scale despite limited dark fiber availability.

\section{Experiment}

\begin{figure*}[t!]
    \centering
    \includegraphics[width=1\linewidth]{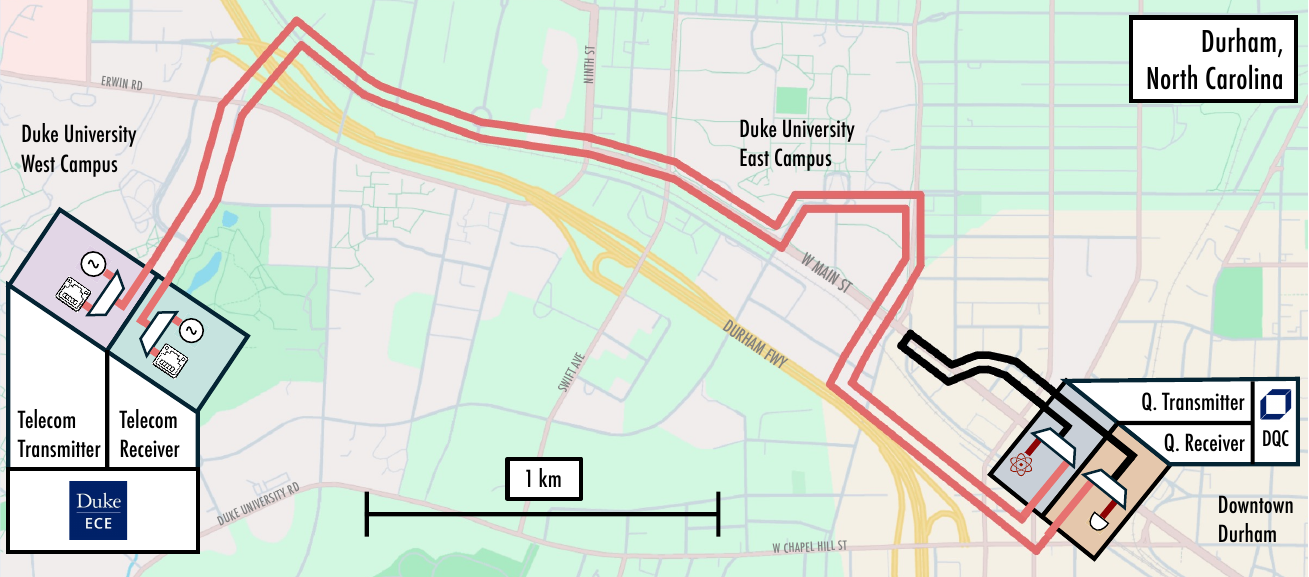}
    \caption{Fiber network map. A quantum transmitter node and quantum receiver node are housed in the Duke Quantum Center. A single \SI{2.8}{\kilo\meter} deployed fiber loop connects the nodes. \SI{1092}{\nano\meter} photons are sent from a trapped ion at the transmitter to photon detectors at the receiver via the fiber loop. Separately, a telecommunication transmitter node located on Duke University West Campus sends Ethernet and 5G signals to the quantum transmitter node where they are wavelength-division multiplexed with the single photons. These co-propagate over the fiber loop. The Ethernet and 5G are demultiplexed from the photons at the quantum receiver node and run to a telecommunication receiver node on West Campus to be decoded.}
    \label{fig:map}
\end{figure*}

\begin{figure*}[t!]
    \centering
    \includegraphics[width=1\textwidth]{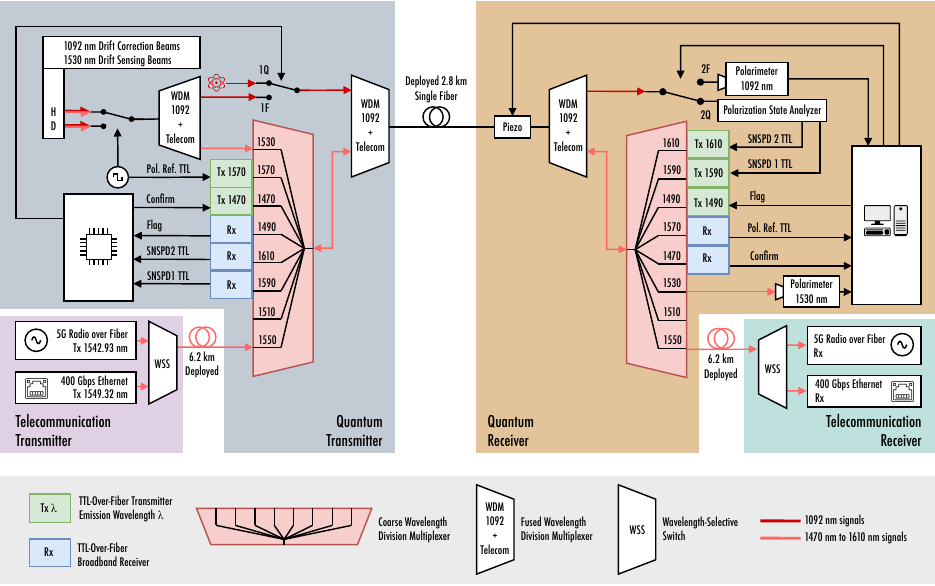}
    \caption{Fiber network layout. The quantum transmitter and quantum receiver are connected via a single deployed fiber. The transmitter hosts the trapped ion, \SI{1530}{\nano\meter} drift sensing beam sources, and \SI{1092}{\nano\meter} drift correction beam sources. The receiver hosts the polarization state analyzer with superconducting nanowire single-photon detectors (SNSPDs), a \SI{1530}{\nano\meter} polarimeter, and a \SI{1092}{\nano\meter} polarimeter. TTL-over-fiber modules are installed in both nodes to facilitate node-to-node communication. During entanglement distribution, SNSPDs detect photons and output TTL pulses which are converted to \SI{1590}{\nano\meter} and \SI{1610}{\nano\meter} optical pulses and sent back to the quantum transmitter over the original fiber link. Independent of the quantum functions, the telecommunication transmitter sends Ethernet and 5G signals over the fiber link to the telecommunication receiver. Optical switches are set to position ``Q'' during quantum entanglement distribution, and position ``F'' during feedback for polarization drift correction.}
    \label{fig:NetworkDiagram}
\end{figure*}

\begin{figure*}[t!]
    \centering
    \includegraphics[width=1\linewidth]{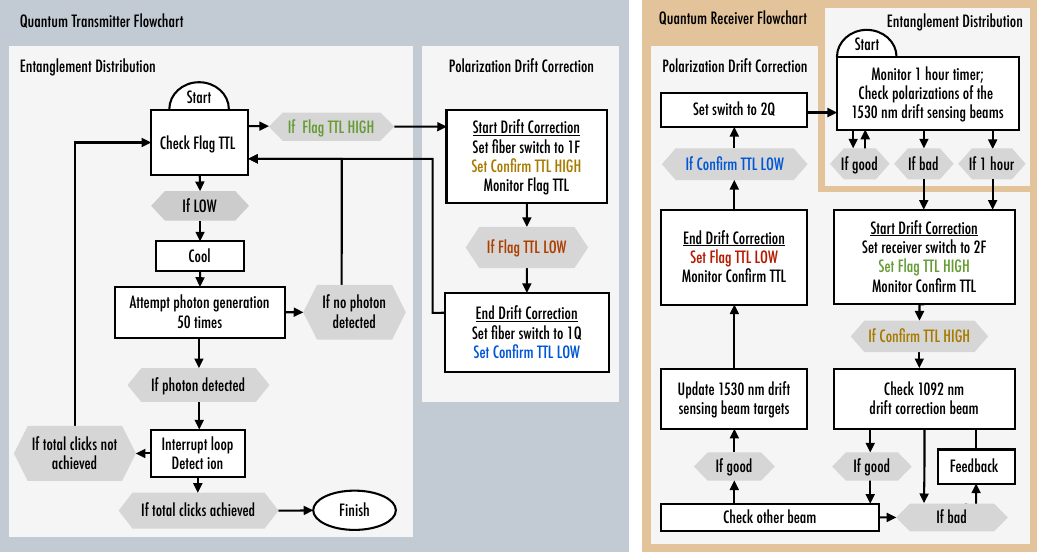}
    \caption{Simplified operation flowchart. The quantum transmitter node control system and receiver node control system work together to execute two modes of operation: entanglement distribution and polarization drift correction. During entanglement distribution, the transmitter regularly checks for fiber instability by reading the Flag TTL. If the fiber is stable, the transmitter attempts to send single photons to the receiver. Detections trigger ion measurement. This proceeds until a user-defined number of total clicks is achieved. During entanglement distribution, the transmitter also sends \SI{1530}{\nano\meter} polarization drift sensing beams to the receiver's polarimeter. The receiver monitors for drift beyond a user-defined threshold. If drift is detected, or if 1 hour elapses without drift correction, the receiver toggles the Flag TTL to alert the transmitter. Both then enter their polarization drift correction loops. The transmitter pauses single photon generation and sends \SI{1092}{\nano\meter} drift correction beams to the receiver. The receiver control system uses a piezoelectric fiber squeezer to return each beam polarization back to its target using gradient descent. Upon completion, the systems re-enter their entanglement distribution loops.}
    \label{fig:flowchart}
\end{figure*}

The experimental apparatus consists of three subsystems: the entanglement distribution system, the polarization drift correction system, and the telecommunication system. Each one has components deployed across multiple network nodes. Figure~\ref{fig:map} shows the four nodes used in the experiment. Quantum transmitter and receiver nodes are located in the Duke Quantum Center (DQC). These nodes are linked by one \SI{2.8}{\kilo\meter} deployed fiber loop. Telecom transmitter and receiver nodes are located on the West Campus of Duke University. The telecom transmitter (receiver) is connected to the quantum transmitter (receiver) by a deployed \SI{6.2}{\kilo\meter} fiber. All deployed fibers run underground.

The entanglement distribution and drift correction systems work in tandem. The former generates ion-photon entangled states at the quantum transmitter node. Photons are sent across the fiber loop to be measured at the quantum receiver node. Simultaneously, the drift correction system propagates \SI{1530}{\nano\meter} light through the fiber loop and monitors the polarization for drift. If drift occurs, the system pauses entanglement distribution and performs polarization correction at \SI{1092}{\nano\meter}. Entanglement distribution resumes upon completion. The telecommunication system runs Ethernet and 5G services over the fiber throughout the experiment. Disruption to entanglement distribution is suppressed with spectral filtering discussed in Section~\ref{subsec:EntanglementDistribution}. The hardware layout in each node is shown in Figure~\ref{fig:NetworkDiagram}. 

\subsection{Entanglement Distribution System}
\label{subsec:EntanglementDistribution}

The entanglement distribution system components are located in the quantum transmitter and receiver nodes. The transmitter houses a four-rod RF electric quadrupole trap which confines a single \textsuperscript{88}Sr\textsuperscript{+} ion. A free-space objective lens with 0.6 numerical aperture collects \SI{1092}{\nano\meter} photons emitted by the ion. The objective is aligned with the quantization axis of the ion, established via magnetic field coils that generate a \SI{4.3}{\gauss} field. Exciting the ion to the $\ket{5P_{1/2}, -1/2}$ state results in decay along the \SI{1092}{\nano\meter} transition to the $4D_{3/2}$ level with $5.6\%$ branching ratio (Figure~\ref{fig:levelDiagram}). $\sigma^{+}$ and $\sigma^{-}$ emissions are coupled into SMF-28 fiber and mapped to horizontal (H) and vertical (V) polarization. This yields the entangled state
\begin{equation}
\label{eqn:en-state-Zbasis}
    \ket{\psi}=\frac{\sqrt{3}}{2}\ket{0}\ket{H}+\frac{1}{2}\ket{1}\ket{V},
\end{equation} 
where $\ket{0}\doteq\ket{4D_{3/2}, -3/2}$ and $\ket{1}\doteq\ket{4D_{3/2}, +1/2}$. 

Once in fiber, photons enter the network shown in Figure~\ref{fig:NetworkDiagram}. They first traverse a micro-electromechanical system (MEMS) switch (Thorlabs OSW22-980E). The switch is set to position 1Q during entanglement distribution. This routes the photons into a fused fiber WDM (Haphit FCLD-0915-2-M305-FC/PC; see Appendix~\ref{app:fusedWDM}) which merges \SI{1092}{\nano\meter} light with telecom-wavelength signals (\SIrange{1470}{1610}{\nano\meter}). The photons are sent over the \SI{2.8}{\kilo\meter} deployed fiber loop (SMF-28). At the receiver, photons enter a piezoelectric fiber squeezer, then are split from the telecom-wavelength signals using another fused fiber WDM. Photons travel into a MEMS switch set to position 2Q during entanglement distribution. They are routed to a polarization state analyzer consisting of a complete set of waveplates (quarter, half, quarter), a Wollaston prism, and two fiber-coupled superconducting nanowire single-photon detectors (SNSPDs). The waveplates are controlled with piezo-actuated rotation mounts. The angles set the measurement basis of the polarization qubit. The total loss of the fiber component chain before the polarization state analyzer is $10.6^{+0.3}_{-0.2}$ dB. Errors are given as maximum and minimum bounds. Individual component losses can be found in Appendix~\ref{app:InsertionLoss}.

Photon detection heralds successful entanglement distribution. The transmitter node must be notified of success. In this work, the transmitter is notified using TTL-over-fiber converters (provided by Optical Zonu Corp.). SNSPD 1 and 2 TTLs are converted into optical pulses at \SI{1590}{\nano\meter} and \SI{1610}{\nano\meter}, respectively. The pulses are sent back through the original fiber to the transmitter node where receivers convert them back to TTL pulses for the control system to read. See Appendix~\ref{app:TTLoF} for more details on the TTL-over-fiber modules which facilitate classical communication between the network nodes. The total delay between ion excitation and reception of SNSPD clicks back at the transmitter node is \SI{28.335}{\micro\second}, well below the ion qubit decoherence time of $>\SI{700}{\micro\second}$~\cite{zalewski2026kilometer}. Measurements of the photon polarization and the ion qubit in several bases are required to evaluate the fidelity of the generated entangled state. We execute ion qubit readout upon any photon detection event, collecting statistics to calculate fidelity.

The quantum transmitter control system runs an entanglement distribution loop during periods when the fiber is stable (Figure~\ref{fig:flowchart}). It checks for fiber stability before entering the loop by reading the Flag TTL value set by the receiver node. If stable, the system enters the entanglement distribution loop. The ion is Doppler cooled with \SI{422}{\nano\meter} and \SI{1092}{\nano\meter} beams (\SI{200}{\micro\second}). The ion is pumped to the $\ket{5S_{1/2},+1/2}$ state (\SI{500}{\nano\second}), then excited to the $\ket{5P_{1/2},-1/2}$ state (\SI{20}{\nano\second}). After the \SI{28.335}{\micro\second} delay, the control system looks for SNSPD TTL pulses indicating successful photon detection. The detection window is \SI{15}{\nano\second}. If no TTL is detected, the loop restarts at the pumping step. If 50 attempts are unsuccessful, the loop restarts at the Flag TTL check. In the event of a photon click, the ion qubit is read out in the Z- or X-basis ($\sim\SI{13}{\milli\second}$). See Appendix~\ref{app:QubitReadout} for readout scheme details. This entanglement distribution loop runs until a user-defined total number of photon clicks is achieved.

During entanglement distribution, telecom signals at \SI{1549.32}{\nano\meter}, \SI{1542.93}{\nano\meter}, \SI{1530}{\nano\meter}, and \SI{1570}{\nano\meter} occupy the deployed fiber. These signals contribute a total optical power of \SI{1.3}{\milli\watt}. The resulting SNSPD background counts are suppressed by three layers of filtering. The quantum receiver fused fiber WDM provides $>10$ dB attenuation. Custom interference filters with optical density $>10$ installed just before the SNSPDs provide additional suppression (AVR Optics). Finally, the Wollaston prism in the polarization state analyzer exhibits large enough chromatic dispersion that telecom light is spatially separated from \SI{1092}{\nano\meter} light at the SNSPD input fibers. These three filtering methods reduce telecom signal-induced SNSPD background counts to dark count level, effectively decoupling photon detection from all telecom traffic implemented in this experiment. 

\subsection{Polarization Drift Correction System}
\label{sec:PolarizationDriftCorrectionSystem}

Every deployed fiber environment exhibits unique mechanical and thermal noise that fluctuates the polarization of transmitted light. When the memory qubit is entangled with photon polarization, such perturbations scramble the state and reduce fidelity. We find the predominant instability affecting the deployed fiber loop to be temperature change arising from day/night transitions causing unitary polarization drifts. Small non-unitary drifts additionally arise due to the multi-mode nature of \SI{1092}{\nano\meter} light in the fiber. See Appendix~\ref{app:FiberStability} for further fiber stability discussion. We implement polarization drift correction to address the unitary drift by periodically pausing entanglement distribution and feeding back on a piezoelectric fiber squeezer. 

Correction is triggered once an hour. We trigger additional corrections as needed if polarization fluctuations are detected. Drift sensing beams at \SI{1530}{\nano\meter} propagate through the fiber during entanglement distribution and are monitored at the quantum receiver. If the polarizations of these \SI{1530}{\nano\meter} beams change beyond a user-defined threshold, correction is triggered.

Drift correction occurs at \SI{1092}{\nano\meter}. The quantum transmitter node houses a source for generating one H- and one D-polarized \SI{1092}{\nano\meter} beam for correction. The source also generates the H- and D-polarized \SI{1530}{\nano\meter} drift sensing beams. A function generator outputs a TTL to an optical switch that selects which polarization propagates through the network. The function generator TTL switches every 1.5 seconds. It is also transmitted to the receiver node via TTL-over-fiber (\SI{1570}{\nano\meter}) such that the receiver control system knows which polarization is being sent (``Polarization Reference TTL'' in Figure~\ref{fig:NetworkDiagram}). During entanglement distribution, the \SI{1092}{\nano\meter} beams are shuttered while the \SI{1530}{\nano\meter} beams enter the switch and a series of multiplexers to merge with the single photons. At the receiver, they are demultiplexed from the photons and read on a polarimeter. 

If correction is triggered, the receiver sets its MEMS switch to position 2F, redirecting \SI{1092}{\nano\meter} light to a polarimeter. It toggles a Flag TTL (\SI{1490}{\nano\meter}) to alert the transmitter of drift. The transmitter node sets its MEMS switch to position 1F and unblocks the \SI{1092}{\nano\meter} drift correction beams. It toggles a Confirm TTL (\SI{1470}{\nano\meter}), indicating to the receiver that polarization correction using the piezoelectric fiber squeezer should commence.

The fiber squeezer acts effectively as a complete set of waveplates for polarization control. During each 1.5-second pulse window, the polarization correction algorithm implements gradient descent to bring the current drift correction beam back to its target, actuating on the four channels of the squeezer. See Appendix~\ref{app:DriftCorrection} for details. Since the four piezos are shared between both drift correction beams and only one beam is measurable at a time, the algorithm alternates between optimizing for the H and D beams on the 1.5-second switching period. Convergence is declared when both beams enter their respective time slots already within the success threshold without requiring correction, verifying that correction of one beam has not disturbed the other.

Upon successful correction, the receiver toggles the Flag TTL. The transmitter sets its switch back to position 1Q and blocks the \SI{1092}{\nano\meter} drift correction beams. The Confirm TTL is toggled to indicate that the transmitter is ready for entanglement distribution. The receiver switch is set back to position 2Q. Entanglement distribution and \SI{1530}{\nano\meter} drift sensing resume.

\subsection{Telecommunication System}

We test the viability of photon coexistence with telecom traffic by merging two types of telecom signals into the quantum link fiber: Ethernet and 5G analog radio-over-fiber.

The Ethernet transmitter consists of a Wistron Galileo Flex-T platform equipped with Lumentum CFP2-DCO coherent pluggable modules running the NEC Network Operating System. This system represents a typical setup used for establishing interconnects between data centers. It realizes 400 Gigabits per second (Gbps) transmission rate by encoding data in the amplitude and phase of a \SI{1549.32}{\nano\meter} carrier beam. Sixteen unique amplitude and phase combinations can be transmitted during a clock cycle (using 16-state quadrature amplitude modulation, or 16-QAM). One fiber supports two such transmission channels as two polarizations are used. Signal decoding occurs at the telecom receiver node. Forward error correction (FEC) is applied to the transmission, utilizing bandwidth overhead to encode redundant bits. 

The 5G New Radio transmitter consists of an RF transceiver (USRP B210) outputting a software-defined waveform to a linear electro-absorption modulator (Optilab LT-12-E-M \SI{12}{\giga\hertz} Lightwave transmitter) which modulates the waveform directly onto a \SI{1542.93}{\nano\meter} optical signal. It establishes a 5G physical downlink shared channel using orthogonal frequency-division multiplexing with 16-QAM on each subcarrier. At the telecom receiver node, the optical signal is detected by an amplified PIN photodiode receiver (Optilab PR-12-B-M \SI{12}{\giga\hertz} PhotoReceiver) and converted back to the RF domain to be received by a second USRP B210. This system represents a typical setup for 5G fronthaul applications in which data is sent from the internet backbone to a client. A central office station, connected to the wider network, pulls the client-requested data and transmits it via analog radio-over-fiber to the base station near the client. The base station decodes the signal and downlinks it to the client through a cell tower. In this work, the telecom transmitter node emulates the central office while the telecom receiver node emulates the base station. To evaluate link performance, error vector magnitude (EVM) of the received 16-QAM constellation is calculated. EVM quantifies the deviation of received constellation points from their ideal positions as a percentage of the root-mean-square (RMS) reference amplitude. Lower values indicate higher signal quality. The 3rd Generation Partnership Project (3GPP) standard mandates EVM $\leq 12.5\%$ for 16-QAM~\cite{3gpp38104}.

At the telecom transmitter node, the Ethernet and 5G signals are multiplexed by a wavelength-selective switch within a reconfigurable optical add-drop multiplexer. They are sent over a \SI{6.2}{\kilo\meter} fiber to the quantum transmitter node, and then multiplexed with the single photons through one of the unused ports of the TTL-over-fiber coarse wavelength-division multiplexer (CWDM) and the fused fiber WDM. After traversing the \SI{2.8}{\kilo\meter} loop, they are demultiplexed at the quantum receiver node using another fused fiber WDM and CWDM. They are sent to the telecom receiver node over another \SI{6.2}{\kilo\meter} fiber to be further demultiplexed and decoded. See Figure~\ref{fig:NetworkDiagram}.

\section{Results}
\label{sec:Results}

\begin{figure*}
    \centering
    \includegraphics[width=0.95\textwidth]{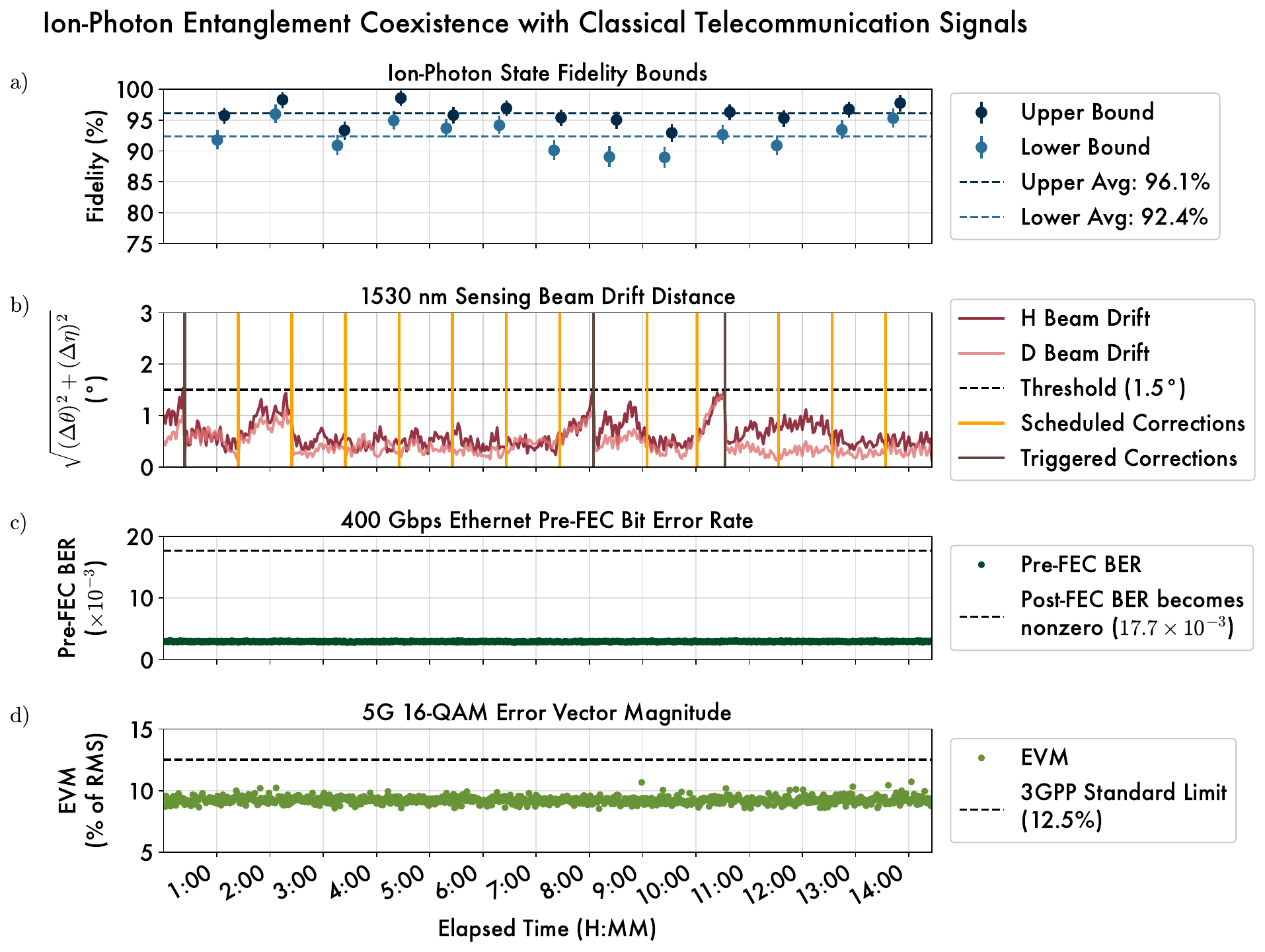}
    \caption{Coexistence demonstration results. (a) Ion-photon entangled state fidelity bounds over 13 hours, 46 minutes, and 9 seconds of operation. (b) \SI{1530}{\nano\meter} H-polarized and D-polarized drift sensing beams co-propagate with the photons during the experiment. Each beam experiences polarization drift from fiber instability. Drift distance $\sqrt{(\Delta\theta)^{2}+(\Delta\eta)^{2}}$ is plotted in degrees. $\Delta\theta$ and $\Delta\eta$ represent the azimuth and ellipticity drift, respectively. Vertical orange lines indicate when scheduled polarization correction occurs. Vertical brown lines indicate when drift-triggered correction occurs. Drift is always calculated as distance from the most recent target polarization, which is updated at the end of every correction. (c) Ethernet co-propagates with the photons. The bit error rate (BER) at the telecom receiver is plotted before forward error correction (FEC) is applied. The dashed line indicates the pre-FEC BER threshold at which error correction no longer succeeds in reducing post-FEC BER to zero. Sufficient power is used such that the pre-FEC BER is well below the threshold, showing that the Ethernet link is functional throughout the experiment. (d) Error Vector Magnitude (EVM) of the 5G 16-state quadrature amplitude modulated signal quantifies the deviation of the real received waveform from the intended waveform. The 3rd Generation Partnership Project defines a maximum EVM threshold of 12.5\%, above which the link breaks. Sufficient transmitter output power is used such that EVM stays well below the threshold throughout the experiment.}
    \label{fig:results}
\end{figure*}

We execute the distributed quantum transmitter/receiver protocol (Figure~\ref{fig:flowchart}) for 13 hours, 46 minutes, and 9 seconds. 32,500 photon detection events are recorded, giving an entanglement generation rate of 39.3 per minute. The measurement basis of the ion-photon entangled state is alternated between ZZ and XX throughout data collection. ZZ measurements produce terms of the bipartite density matrix $\rho_{0 H, 0 H}$, $\rho_{0 V, 0 V}$, $\rho_{1 H, 1 H}$, and $\rho_{1 V, 1 V}$. XX measurements produce terms of a rotated density matrix $\tilde{\rho}_{0 H, 0 H}$, $\tilde{\rho}_{0 V, 0 V}$, $\tilde{\rho}_{1 H, 1 H}$, and $\tilde{\rho}_{1 V, 1 V}$. The upper bound $\mathcal{F}_{+}$ and lower bound $\mathcal{F}_{-}$ of fidelity,
\begin{equation}
\begin{aligned}
    \mathcal{F}_{\pm} & = \frac{3}{4} \rho_{0 H, 0 H} + \frac{1}{4} \rho_{1 V, 1 V}  + \frac{\sqrt{3}}{4} (\tilde{\rho}_{1 V, 1 V} +\tilde{\rho}_{0 H, 0 H} \\ & - \tilde{\rho}_{0 V, 0 V} - \tilde{\rho}_{1 H, 1 H}  \pm 2 \sqrt{\rho_{0 V, 0 V}  \rho_{1 H, 1 H}} ),
\end{aligned}
\end{equation}
can then be calculated on a rolling basis over the experiment~\cite{AnaTimeBin}. The time series is shown in Figure~\ref{fig:results}(a). It demonstrates high-fidelity performance throughout the data collection period while all telecom signals are present, with average fidelity bounds of $92.4(4)\%\le\mathcal{F}\le 96.1(4)\%$. Uncertainties are estimated using nonparametric bootstrapping with replacement, and show confidence intervals incorporating 68\% of the bootstrapped outcomes. Magnetic field variations in the lab contribute to qubit splitting instability on the scale of \SI{1.5}{\kilo\hertz} during the experiment. Single-qubit rotation errors then occur and imprint on ion X-basis measurements. This is one contributor to small fidelity bound fluctuations in Figure~\ref{fig:results}(a). Such effects can be mitigated using standard magnetic field stabilization techniques~\cite{ruster2016long}. Non-unitary polarization fluctuations arising from the multi-modality of \SI{1092}{\nano\meter} light in the deployed fiber additionally contribute to fidelity fluctuations. See Appendix~\ref{app:FiberStability} for further discussion. The remaining error sources for this system are detailed in Ref.~\cite{zalewski2026kilometer}. 

The H- and D-polarized \SI{1530}{\nano\meter} sensing beams are monitored during entanglement distribution. Their drift distances are plotted in Figure~\ref{fig:results}(b). See Appendix~\ref{app:DriftCorrection} for details on drift calculation. A trigger threshold distance of $1.5^{\circ}$ is used; any polarization excursion exceeding this threshold triggers drift correction. $1.5^{\circ}$ is chosen as it is well below the level at which significant infidelity arises, while being enough above the polarimeter measurement noise that spurious triggers do not occur.

Correction is executed fifteen times during the experiment. Three corrections are triggered by real fiber drifts, while the remainder are scheduled hourly corrections. The average duration of correction is 24.7 seconds. The total time spent in drift correction mode over the entire experiment is 6 minutes and 11 seconds. The system executes entanglement distribution mode for the remaining 13 hours, 39 minutes, and 58 seconds, occupying $99.3\%$ of experiment time. 

400 Gbps Ethernet at \SI{1549.32}{\nano\meter} runs over the fiber during data collection. We measure the signal's pre-forward error correction bit error rate (pre-FEC BER) at the telecom receiver node, shown in Figure~\ref{fig:results}(c). The output power of the transmitter is set to minimize pre-FEC BER. We show that under this condition, the pre-FEC BER is low enough that error correction reduces the residual BER to zero. This demonstrates that the Ethernet link is error-free and functional during entanglement distribution.

The 5G 16-QAM signal also runs over the fiber during data collection. We again set the output power to minimize error, yielding an average EVM of $\sim$9\% as shown in Figure~\ref{fig:results}(d). This lies well below the 3GPP standard limit of 12.5\%, demonstrating that the 5G link remains fully functional during entanglement distribution.

The total telecom optical power traversing the fiber is \SI{1.3}{\milli\watt}, measured just before the receiver node's first WDM. This power measurement includes the Ethernet, 5G, \SI{1530}{\nano\meter} fiber sensing, and \SI{1570}{\nano\meter} polarization reference TTL signals (see Appendix~\ref{app:TelecomPower} for further Ethernet and 5G details). None of these signals are shuttered during photon detection windows. Fused fiber WDMs, high optical density filters, and spatial filtering from chromatic dispersion in the polarization state analyzer Wollaston prism are sufficient to reduce the background counts to negligible levels. This ensures that the ion-photon state fidelity suffers no impact from coexistence.

\section{Outlook and Future Work}

Practical, city-scale quantum networking places considerable demands on distributed control, polarization stabilization, and high-fidelity sources of low-loss photons. We introduce an entanglement distribution system which integrates all of these capabilities into a deployable quantum transmitter/receiver package. In the experiment, each site offers wavelength-division multiplexing ports which permit telecom signals to pass through the fiber with single photons at no cost to entanglement fidelity. Our results show that coexistence between memory-based quantum networking and classical networking is achievable with standard components. This is made possible by using low-loss photon wavelengths outside telecom bands. Such capability condenses quantum link fiber requirements by employing one fiber for all quantum and classical utilities. With this approach, quantum link deployment over fiber networks crowded with telecom traffic becomes possible.

Our spectral filtering suppresses the telecom-induced photon detector background to dark count level. This indicates that significant additional classical bandwidth is still available for multiplexing more telecom traffic onto the fiber before infidelity arises.

The current entanglement rate of 39.3 per minute is limited by the low branching ratio of the \SI{1092}{\nano\meter} transition, finite numerical aperture of the collection optics, and photon detection wait time. Cavity integration and ion multiplexing can be used to boost the rate~\cite{BenMultiplexing,DuanMultiplexing}.

The fiber loop in this work inhabits a stable underground tunnel which is amenable to polarization encoding, representative of many conduit environments found in cities. High fidelity is achieved despite the multi-mode behavior of \SI{1092}{\nano\meter} photons in fiber. Aerial fibers can present polarization perturbations on faster time scales and strengths than those encountered here~\cite{Kucera14,BARQNET}. The implemented polarization drift correction handles such instability at the cost of reduced quantum link uptime. When this tradeoff becomes unfavorable, alternative photon encoding schemes can be considered. The time-bin basis offers intrinsic fidelity robustness to perturbations in highly unstable fibers at the expense of entanglement rate~\cite{AnaTimeBin,EschnerTimeBin,Covey,SagnikTimeBin}. The WDMs, MEMS switches, and spectral filtering components used in this work are agnostic to the photon basis, so isolation of quantum and classical operations can be maintained regardless of encoding scheme. The presented quantum transmitter/receiver package unlocks entanglement distribution for real-world quantum networks in a multitude of environments.

\begin{acknowledgments}

We thank Michael Straus, Isabella Goetting, and Ashish Kalakuntla for helpful discussions.
Johnny Bell and Dickson Clifford in the Duke Office of Information Technology furnished the deployed fiber links used in this work. We additionally thank Britt McGinn of Optical Zonu Corp. for assistance with TTL-over-fiber capabilities.
This work was supported by the Army Research Office (grants W911NF-19-10296, W911NF-17-S-0002-0, W911NF-19-20181, and W911NF-22-10032), the National Science Foundation Convergence Accelerator program (OIA-2134891), and the Software-Tailored Architecture for Quantum CoDesign (STAQ) Award (PHY-2325080), and awards CNS-2211944, CNS-2330333, CNS-2443137, and CNS-2450567, as well as funding from Duke University under the Beyond-the-Horizon and DST-Launch initiatives.

\end{acknowledgments}

\bibliography{apssamp}

\appendix

\section{Fiber Network Components}

\subsection{Fused Fiber Wavelength-Division Multiplexers}
\label{app:fusedWDM}

We use custom fused fiber WDMs (Haphit FCLD-0915-2-M305-FC/PC) to split and merge \SI{1092}{\nano\meter} photons and telecom traffic between \SI{1470}{\nano\meter} and \SI{1610}{\nano\meter}. The fused fiber WDMs are fabricated to optimally multiplex \SI{1092}{\nano\meter} and \SI{1550}{\nano\meter} light. We therefore characterize the insertion loss of a fused fiber WDM across the out-of-specification wavelengths. Data is shown in Table~\ref{tab:FusedWDMLoss}. 

The WDM has one common input port and two output ports. One of the outputs transmits telecom light while blocking \SI{1092}{\nano\meter} light. We find that insertion losses for wavelengths at the edges of the \SI{1470}{\nano\meter} to \SI{1610}{\nano\meter} band are higher than for those in the center of the band, but easily surmounted by the TTL-over-fiber transmitter powers. The second output port blocks telecom light and transmits \SI{1092}{\nano\meter} light. We find the insertion loss through this port is maximal at \SI{1550}{\nano\meter} as expected, and reduces for wavelengths away from \SI{1550}{\nano\meter}. This leads to extra power leakage into our polarization state analyzer. This leakage would reduce fidelity, so we additionally insert interference filters as well as spatially filter the telecom light, reducing detector background to negligible levels.

\begin{table}
    \centering
    \caption{Fused WDM insertion loss. Pass port data refers to the loss measured between the common input port and the telecom output port. Block port data refers to the loss measured between the common input port and the \SI{1092}{\nano\meter} output port, which is meant to block telecom light.}
    \label{tab:FusedWDMLoss}
    {\renewcommand{\arraystretch}{1.3}
    \begin{tabular}{ccc}
        \toprule
        Wavelength (nm) & Pass Port Loss (dB) & Block Port Loss (dB) \\
        \midrule
        1470 & 1.01 & 12.6 \\
        1490 & 0.42 & 14.6 \\
        1510 & 0.32 & 15.6 \\
        1530 & 0.22 & 22.9 \\
        1550 & 0.27 & 36.9 \\
        1570 & 0.25 & 25.2 \\
        1590 & 0.30 & 17.5 \\
        1610 & 0.45 & 13.5 \\
        \bottomrule
    \end{tabular}}
\end{table}

\subsection{TTL-Over-Fiber Modules and Coarse Wavelength-Division Multiplexer}
\label{app:TTLoF}

The Optical Zonu TTL-over-fiber package facilitates classical TTL communication between the quantum transmitter and receiver nodes. The TTL transmitters convert TTL pulses to optical pulses with latency $< \SI{7}{\nano\second}$, while the TTL receivers convert them back with latency $< \SI{6}{\nano\second}$. Each transmitter outputs 3 dBm optical pulses at one of eight wavelengths in the coarse wavelength-division multiplexing standard. The available wavelengths are \SI{1470}{\nano\meter}, \SI{1490}{\nano\meter}, \SI{1510}{\nano\meter}, \SI{1530}{\nano\meter}, \SI{1550}{\nano\meter}, \SI{1570}{\nano\meter}, \SI{1590}{\nano\meter}, and \SI{1610}{\nano\meter}. The package includes a bidirectional coarse wavelength-division multiplexer (CWDM) for each network node. Not all wavelength channels are used in the experiment. The quantum transmitter node requires only two TTL transmitters, and the quantum receiver node only requires three. The receivers that accompany each transmitter are broadband and can be used with any transmitter wavelength to convert optical pulses back to copper signals. 

Empty channels on the coarse wavelength-division multiplexers are thus used to multiplex the \SI{1530}{\nano\meter} drift sensing beams, \SI{1549.32}{\nano\meter} Ethernet, and \SI{1542.93}{\nano\meter} 5G signals with the single photons. While the wavelengths of the Ethernet and 5G do not exactly match the wavelength of the \SI{1550}{\nano\meter} CWDM channel used, the CWDM still merges the signals, albeit with additional loss. The loss is easily surmounted given the ample transmission powers available in the telecom band.

\section{Fiber Component Losses}
\label{app:InsertionLoss}

The insertion losses of all fiber components in the \SI{1092}{\nano\meter} photon path are shown in Table~\ref{tab:InsertionLoss}. Errors are given as maximum and minimum bounds. 

\begin{table}
    \centering
    \caption{Insertion loss of individual fiber network components. Error bars indicate hard upper and lower bound measurements. Discrepancy between total loss and sum of individual losses arises from fiber mating efficiency variation.}
    \label{tab:InsertionLoss}
    \renewcommand{\arraystretch}{1.5}
    \begin{tabular}{lc}
        \toprule
        Component & Loss (dB) \\
        \midrule
        Transmitter MEMS switch        & $2.58^{+0.08}_{-0.08}$ \\
        Quantum transmitter WDM         & $0.74^{+0.19}_{-0.14}$ \\
        Fiber loop                      & $3.32^{+0.12}_{-0.10}$ \\
        Fiber squeezer                  & $1.74^{+0.09}_{-0.08}$ \\
        Quantum receiver WDM            & $0.45^{+0.18}_{-0.11}$ \\
        Receiver MEMS switch           & $2.60^{+0.09}_{-0.07}$ \\
        Total Fiber Component Chain     & $10.6^{+0.3}_{-0.2}$ \\
        \bottomrule
    \end{tabular}
\end{table}

\section{Ion Qubit Readout}
\label{app:QubitReadout}

We implement ion qubit readout using a three-step process. After photon detection, the ion qubit is rotated using \SI{1004}{\nano\meter} co-propagating Raman beams to set the measurement basis. The pulse takes \SIrange{0}{76.3}{\micro\second}, depending on the basis selected. Population in the $\ket{4D_{3/2},+1/2}$ state is then shelved to the $\ket{4D_{5/2}}$ level using resonant \SI{1004}{\nano\meter} $\sigma^{-}$ and resonant \SI{408}{\nano\meter} $\sigma^{+}$ beams for \SI{180}{\micro\second}. The remaining population in the $\ket{4D_{3/2},-3/2}$ state is read out using fluorescence detection with \SI{422}{\nano\meter} and \SI{1092}{\nano\meter} beams for \SI{13}{\milli\second}.

\section{Deployed Fiber Loop Stability}
\label{app:FiberStability}

\begin{figure}
    \centering
    \includegraphics[width=1\linewidth]{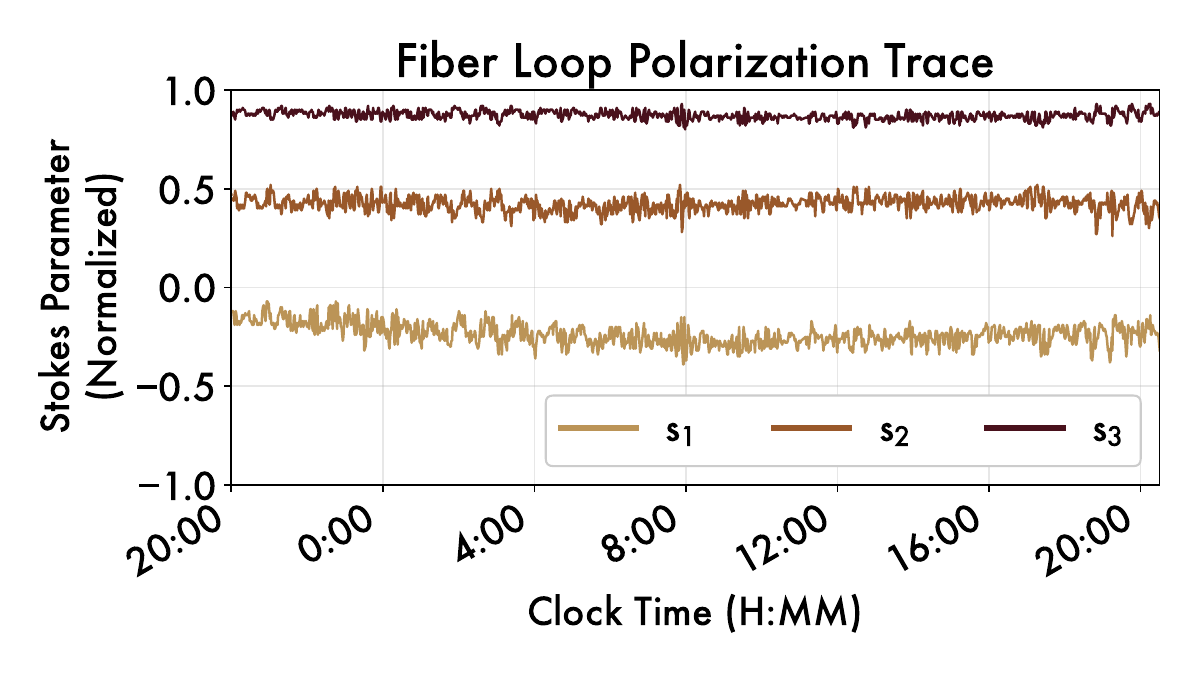}
    \caption{Stokes parameters of a classical \SI{1092}{\nano\meter} beam traversing the fiber loop. Slow drift occurs over the day due to temperature drift. Noise on the time scale of a minute arises due to small multi-mode effects in the fiber.}
    \label{fig:Stokes}
\end{figure}

We study the stability of the deployed fiber in this work by sending \SI{1092}{\nano\meter} classical light through the loop. Figure~\ref{fig:Stokes} shows the Stokes parameters of the output beam over 24 hours. We find fluctuations on two distinct time scales. A slow drift on the time scale of hours arises due to the changing temperature of the day. A faster noise on the time scale of minutes is superimposed on the drift. APC fiber tips do not eliminate the fluctuation, showing that it is not fundamentally an etalon effect. Additionally, this fast noise does not appear on \SI{1550}{\nano\meter} classical light sent through the fiber, indicating wavelength dependence. 

The effect arises from the multi-mode nature of \SI{1092}{\nano\meter} light within the fiber. \SI{1092}{\nano\meter} lies below the single-mode cutoff wavelength of the standard SMF-28 fiber used in the loop. Two spatial modes are permitted at this wavelength~\cite{jung2009comparative}: the fundamental LP\textsubscript{01} mode found in conventional single-mode operation, and the two-lobed LP\textsubscript{11} mode which exhibits cylindrical asymmetry. By using \SI{1092}{\nano\meter}, we operate in the ``few-mode fiber'' regime~\cite{FewModeFibers,liu2018intrinsic,yoda2006beam}. Power predominantly occupies the fundamental mode, but a small portion occupies the higher-order mode. Several mechanisms can cause light to pass between these modes: manufacturing defects that lead to core non-circularity, symmetry-breaking bends along the path, and refractive index irregularities generating inter-modal leakage~\cite{kutz1998mode}.

Each mode has a distinct index of refraction. This gives rise to differential mode delay and mode-dependent polarization transformation when traversing the fiber. Furthermore, the refractive indices experience differing sensitivities to temperature change. Because the output of fiber is the waveform sum of the outgoing modes, these differential mode behaviors will be imprinted as interference effects on the light. Small temperature drifts differentially transforming the polarization of each spatial mode cause the waveform sum of the outgoing light to exhibit unstable total polarization. 

\subsection{Implications for Entanglement Distribution}

When sending single photons, the ideal behavior would be maintaining all light in the LP\textsubscript{01} mode, with the polarization states retaining their orthogonality across the entire transmission. If asymmetry induces preferential leakage of one polarization into the higher-order mode, and that mode experiences a polarization transformation differing from the LP\textsubscript{01} component, the orthogonality of polarization encoding will not be perfectly preserved. Moreover, loss from micro-bends and splices can be mode-dependent, offering a mechanism for polarization-dependent loss to occur. In a classical beam, this manifests as output polarization drift correlated with power fluctuation. In an ion-photon entangled state, this reduces fidelity.

While polarization drift correction can revert \SI{1092}{\nano\meter} polarizations to any user-defined target, the feedback implemented in this work does not address polarization orthogonality reduction or polarization-dependent loss because they are non-unitary in nature. These multi-mode effects partially contribute to the fidelity bound fluctuations present in Figure~\ref{fig:results}(a). 

The degree of multi-modality in the deployed fiber is nonetheless sufficiently small that high fidelity is still achieved. Many underground conduits offer similar stability, adequate for high-fidelity entanglement distribution despite multi-mode behavior. However, more research is needed to understand how multi-modality affects polarization-based entanglement distribution in fibers with aerial segments and highly unstable temperatures.

Finally, time-bin encoding offers a route to stable fidelity in the case of severe multi-modality. Provided the time bins are matched in polarization, the time-bin basis gives intrinsic fidelity robustness to polarization transformations, even when non-unitary in nature. The tradeoff for fidelity robustness is entanglement rate reduction.

\subsection{Implications for Polarization Drift Correction}

A hallmark of polarization-dependent loss in this experiment is time variation of the angle between the H and D \SI{1092}{\nano\meter} drift correction beams. In the absence of multi-mode effects, these polarizations form a $90^\circ$ angle on the Poincar\'{e} sphere which should be constant regardless of rotations due to fiber drift. With the polarization-dependent loss non-unitary, we measure a drifting angle between these polarizations on the one-minute time scale. Compensation for this effect is implemented each time drift correction occurs by measuring the angle between these beams and updating the target polarizations to account for this relative angle drift. Discrepancy from $90^\circ$ never exceeds $5^\circ$ during the experiment, indicating the small scale of this effect.

\section{Polarization Drift Correction Details}
\label{app:DriftCorrection}

During drift correction, the quantum transmitter node sends the H- and D-polarized \SI{1092}{\nano\meter} drift correction beams one at a time, alternating every 1.5 seconds. The quantum receiver system implements gradient descent to bring the current \SI{1092}{\nano\meter} drift correction beam to its target. In every descent iteration, the algorithm varies the control voltage of each piezoelectric fiber squeezer channel and measures the angular error $\epsilon = \sqrt{(\Delta\theta)^2 + (\Delta\eta)^2}$ between the measured polarization azimuth and ellipticity ($\theta,\eta$) and the target polarization azimuth and ellipticity ($\theta_{0},\eta_{0}$). The channel voltages are then updated in the direction that reduces $\epsilon$. Step sizes scale with $\epsilon$. Descent is executed until the end of each 1.5-second window.

Drift correction can be triggered if the polarization of any \SI{1530}{\nano\meter} drift sensing beam changes beyond a user-defined threshold. The polarization transformation applied by gradient descent specifically reverts the \SI{1092}{\nano\meter} drift correction beams to their target polarizations. The transformation applied by the fiber squeezer is wavelength-dependent. Consequently, the transformation at \SI{1530}{\nano\meter} will not necessarily revert the \SI{1530}{\nano\meter} drift sensing beams back to their pre-drift polarizations. However, their utility for sensing lies not in their absolute polarizations, but in their polarization changes induced by fiber instability. Any new, post-correction \SI{1530}{\nano\meter} polarizations serve equally well as targets from which to calculate drift distances. Therefore, the drift sensing target polarizations are updated at the end of every correction to be the new, current \SI{1530}{\nano\meter} beam parameters. Drift distance plotted in Figure~\ref{fig:results}(b) reports the distance from the most recent drift sensing target, updated during the most recent drift correction.

\section{Effect of Ethernet and 5G on Fidelity}
\label{app:TelecomPower}

\begin{figure}
    \centering
    \includegraphics[width=1\linewidth]{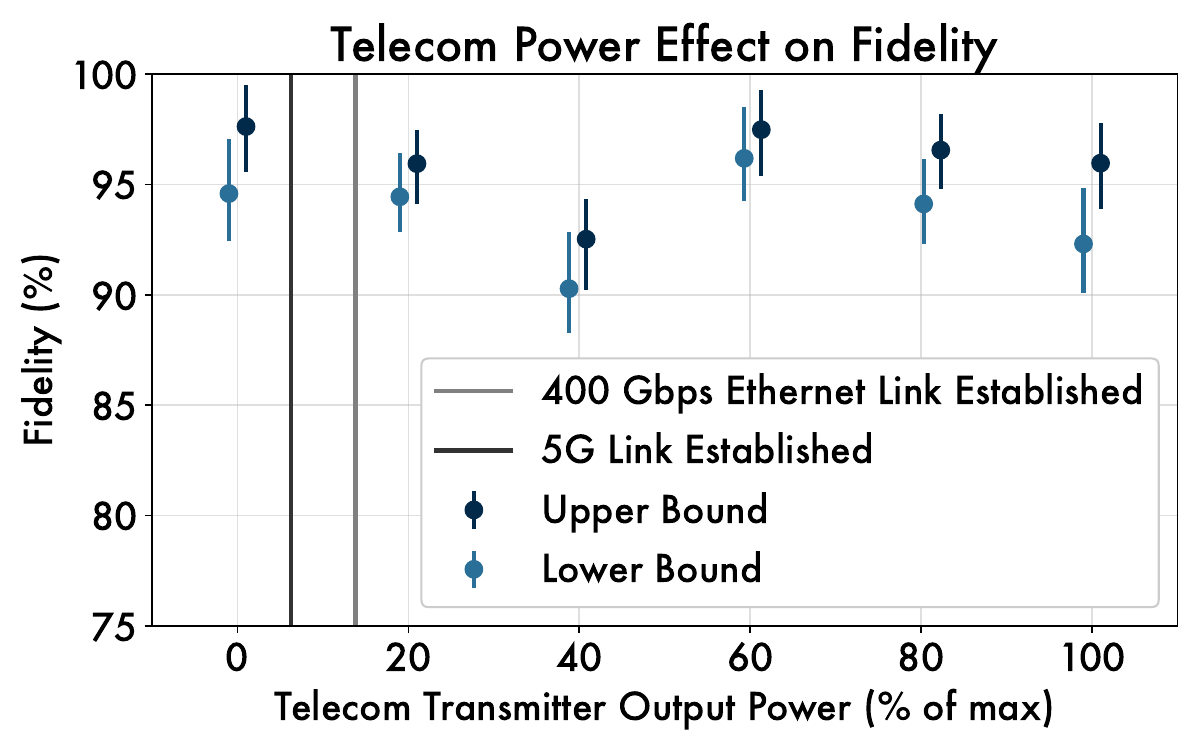}
    \caption{Effect of telecom signals on fidelity. Vertical lines indicate the minimum transmission power required to establish a link.}
    \label{fig:TelecomFidelity}
\end{figure}

We characterize the effect of the Ethernet and 5G signals on ion-photon entangled state fidelity by measuring fidelity at varying telecom transmission powers. Figure~\ref{fig:TelecomFidelity} shows fidelity bounds as a function of telecom transmission power. Ethernet and 5G transmitters are ramped from minimum to maximum output in tandem. The fidelity remains high across all operating powers. During the experiment presented in the main text, both Ethernet and 5G transmitters are set to maximum output power in order to minimize transmission error, amounting to a total of $\sim\SI{0.1}{\milli\watt}$. Given that detector background remains minimal across the telecom transmission powers, additional telecom traffic could still be added before observing fidelity reduction. The threshold at which infidelity arises is not yet reached.

\end{document}